# Ballistic-diffusive model for heat transport in superlattices and the minimum effective heat conductivity

Federico Vázquez 1, Péter Ván 2,3,4,* and Róbert Kovács 2,3,4

1      UAEM, Science Research Center, Dep. of Physics, 62209 Cuernavaca, Av. Universidad 1001, México; vazquez@uaem.mx

2      BME, Faculty of Mechanical Engineering, Dep. of Energy Engineering, 1111 Budapest, Műegyetem rkp. 2, Hungary;

3   MTA Wigner Research Centre for Physics, Dep. of Theoretical Physics, 1121 Budapest, Konkoly-Thege M. út 29-33. Hungary;

4   Montavid Thermodynamic Research Group, Hungary;

*      Correspondence: van.peter@wigner.mta.hu.

Abstract: There has been much interest in semiconductor superlattices because of showing very low thermal conductivities. This makes them especially suitable for applications in a variety of devices for thermoelectric generation of energy, heat control at the nanometric length scale, etc. Recent experiments have confirmed that the effective thermal conductivity of superlattices at room temperature have a minimum for very short periods (in the order of nanometers) as some kinetic calculations had anticipated previously. This work will show advances on a thermodynamic theory of heat transport in nanometric 1D multilayer systems by considering the separation of ballistic and diffusive heat fluxes, which are both described by Guyer-Krumhansl constitutive equations. The dispersion relations, as derived from the ballistic and diffusive heat transport equations, are used to derive an effective heat conductivity of the superlattice and to explain the minimum of the effective thermal conductivity.

## 1. Introduction

This work was motivated by the increasing interest in heat transport in superlattices due to their potential applications in a variety of devices, e.g. for the thermoelectric generation of energy, the design of intelligent coatings for temperature conditioning of enclosures, the construction of devices for processing information, the storage of thermal energy, design of biomedical devices and others [1]. Although the applications of superlattices started several years ago, the nature of the heat transport regime that takes place in them is still being discussed. Particularly, the description of the effects of wave properties of phonons when the film thickness approaches the mean attenuation length of heat carriers continues receiving theoretical and experimental attention [9-13,23] .

The heat transport at the nanoscale deviates from that predicted by the Fourier law, which is able to describe only diffusive transport . When the thickness of a heat conducting layer approaches the nanoscale, other non-Fourier types of transport appear, such as ballistic and wave-like [2-4]. The first arises when the heat carriers move without dispersion between the walls of the material and do not interact with the bulk. The second is a result of the coherent coupling of heat carriers. It arises when there are no dispersive processes that eliminate their phase information.

The diffusive transport and wave propagation (ballistic / second sound) had been experimentally detected only in cold crystals, liquid He, etc., where both heat transport regimes coexist [5-8]. However, since several years ago, some experimental groups have measured the effective thermal conductivity of superlattices [9-12] finding clues to the wave properties of phonons in superlattices at environment temperature. Both transport regimes have also been studied in nanostructured heat sources [13].

Saha and collaborators [9] in epitaxial metal / semiconductor superlattices TiN / (Al, Sc) N with constant thickness increased the density of periods and detected a minimum in the thermal conductivity in thicknesses of the order of the wavelength of the heat carriers. The authors used thermo-reflection in the time domain (TDTR) to measure the thermal conductivity of the samples of a TiN / AlScN superlattice. From the wave point of view the explanation is simple [9]: "For very short periods, when the wavelengths of the dominant phonons involved in heat transport become comparable with the thickness of the period, the individual layers lose their identity and the whole material behaves as an effective medium where the waveforms of the phonon propagate without dispersion in the interfaces, increasing the thermal conductivity." The results by Saha et al. showed a clear minimum (at 4 nm of thickness of period) in the thermal conductivity. In shorter period thicknesses (less than 4 nm), the thermal conductivity increased when the thickness of the period decreased since the phononic modes that do not disperse in the interface contribute significantly to thermal transport. The findings of Saha's group reinforced those obtained some years before by Ravichandran [11] on the existence of a minimum of the effective heat conductivity in superlattices.

In another experimental approach, in measurements at room temperature of effective thermal conductivity as a function of the number of periods in GaAs / AlAs superlattices, keeping constant the thickness of the period [12], it was also found the wave regime of heat propagation. Each internal interface scatters the phonons in a diffuse way, eliminating their phase information. Then the interface behaves as a thermal resistance in such a way that many equivalent interfaces in the series lead to an effective thermal conductivity, in the direction perpendicular to the interfaces, which is approximately independent of the number of layers. However, if the phononic phase is conserved in the interfaces of the superlattice and if the anharmonic dispersion is minimal, the superposition of the Bloch waves creates stop bands and modifies the phononic band structure. In this regime, the mean free paths of the phonons (MFP) are equal to the length of the sample, which leads to a thermal conductivity that is linearly proportional to the total thickness of the superlattice. In addition, the MFP of the phonons is equal to the length of the sample, which leads to a thermal conductivity linearly proportional to the total thickness of the superlattice. The authors carried out measurements of thermal conductivity by time domain thermo-reflection (TDTR) in samples at temperatures between 30 and 300 K. Below 150 K, linearity of thermal conductivity versus length (number of periods) revealed that heat conduction was of the wave type.

Some microscopic theories about heat transport, based on the Boltzmann equation, treat heat carriers (phonons and electrons) as particles but ignores their wave properties. All of them predict that the thermal conductivity perpendicular to the layers in a superlattice decreases as the space between layers is reduced. But, as it may be concluded from the experiments mentioned above, as the thickness of the period becomes smaller than the wavelength of the phonon, the thermal conductivity increases. This disagreement is solved by calculations that include the MFP of the phonons. For thinner layers than the MFP, wave theory is applied. While for layers thicker than the MFP, the theory of particles is applied. The combined theory predicts a minimum of thermal conductivity for a certain value of thickness which depends on the average MFP, and therefore on the temperature [14].

In Simkin [15] formulation of oscillator chains an imaginary part of the form ( is the phononic MFP) is added to the wave number to obtain a minimum in the effective thermal conductivity accordingly with Saha's experiments. The argument used by Simkin is striking: "Phenomenologically, an imaginary part of the wave number of the form can be introduced for …" No other justification is given. Macroscopically, the effective thermal conductivity seems to be determined solely by the steady state which depends on the boundary conditions, the transport coefficients and the interfacial thermal resistance.

The macroscopic models of heat transfer can predict the general form of heat propagation and also can connect the different levels of the hierarchical material behavior [16]. The hyperbolic models with relaxation type dissipation predict a wave behavior of heat but they are not able to incorporate emerging processes related to the coherent behavior of phonons and also the quantitative modelling of ballistic propagation can be problematic [17]. These aspects of the dynamics of non Fourier heat propagation and in particular the form of the evolution equations determines the stationary processes, as well. The analysis using Fourier propagation modes can shed light on the interaction of heat waves with the microstructure. It can be shown that the long wavelength modes have an attenuation length that tends to infinity when the wave number tends to zero. The larger the wavelength the larger the penetration of the wave in the material, in such a way that the long wavelength modes do not interact with the interfaces. If we add to this the fact that long wavelength modes transport most of the heat at the nanoscale it should be possible to explain the behavior of thermal conductivity in superlattices. These issues were addressed in this work.

**2. Methods**

   2.1 Two temperature Ballistic-Diffusive model

   We propose to use irreversible thermodynamics with two vectorial dissipative fluxes (internal variables) to incorporate features of the wave-like behavior of the heat carriers and to describe transient processes associated with the results reported by the experiments above mentioned on superlattices. The theory of internal variables is easily applicable to the multilayer system and we hope that it open new possibilities for analysis.

   Ballistic-diffusive models for heat transport are based on the assumption that there exist two kinds of heat carriers in the system. Those which propagate diffusively that are scattered in the core of the material, and those which propagate ballisticaly interacting only with the boundaries or the interfaces of the system. This assumption has also been made in other systems where there are two or more types of heat carriers. For example, in thermoelectric materials where it is assumed that heat is transported by electrons and phonons. Also in polar semiconductors in which there is electron and hole transport as well as phonons. From a microscopic point of view, the ballistic-diffusive models have been obtained from the Boltzmann equation [18]. The idea is to divide the distribution function into two parts, namely, one that describes the phonons that are generated at the boundaries of the material and that travel through it without being scattered (ballistic), and another one that describes the phonons being scattered in the bulk (diffusive). Chen showed that, up to first order in the expansion of the diffusive distribution function in terms of spherical harmonics, the associated heat flux can be described by an equation of the Cattaneo type, while the part of the heat flux transported by the ballistic phonons is obtained as a solution of the Boltzmann equation. The Cattaneo equation for the diffusive flow is coupled with the ballistic heat flow, so that the solution procedure involves first solving the Boltzmann equation to obtain the ballistic flux and then introducing it into the Cattaneo equation to obtain the diffusive heat flux.

In thermodynamic theories of heat transport, ballistic-diffusive models are obtained assuming that the heat flow can be separated into the ballistic and diffusive components. It has been then assumed that the diffusive part is described by the Cattaneo equation and the ballistic part by Guyer-Krumhansl equation [19,20]. We made the basic assumption that there exist two types of heat carriers and that the heat flux is separable obtaining the constituent equations for the ballistic and diffusive heat flux from the second law of thermodynamics. This was done by including in the space of variables the ballistic and the diffusive heat fluxes. The system consists of two particle populations, namely, ballistic and diffusive heat carriers which have inner energies $e_b$ and $e_d$, and heat fluxes $q_b$ and $q_d$, respectively. In this way, the variable space becomes $\{e_b, e_d, q_b, q_d\}$. Then the balances of the component energies are:

$$C_b \frac{\partial T_b}{\partial t} + \frac{\partial q_b}{\partial x} = -Q - AT_b, \qquad (1)$$

$$C_d \frac{\partial T_d}{\partial t} + \frac{\partial q_d}{\partial x} = Q + AT_d, \qquad (2)$$

where use has been made of the caloric equations $e_d = C_d T_d$ and $e_b = C_b T_b$, being $T_d$ and $T_b$ interpreted as quasi-temperatures and $C_d$ and $C_b$ specific heats of the species. We have assumed that the heat exchange between the ballistic and diffusive parts decomposes to a dissipative part $Q$ and a non-dissipative one that is linear in the quasi-temperature. $A$ is a constant quantity Note that Eqs. (1) and (2) are not counterbalanced which implies that the total internal energy balance equation has a source term. One can see that Eqs. (1) and (2) are more general than the corresponding Chen's internal energy balance equations [18]. We can obtain Chen's equations from Eqs. (1) and (2) by assuming that $Q = 0$ and $A = \dfrac{C_b}{\tau_b}$. In one dimension we obtain

$$C_b \frac{\partial T_b}{\partial t} + \frac{\partial q_b}{\partial x} = -\frac{C_b T_b}{\tau_b}.$$

$$C_d \frac{\partial T_d}{\partial t} + \frac{\partial q_d}{\partial x} = \frac{C_b T_d}{\tau_b},$$

Chen's equations are obtained approximately if $T_d \approx T_b$. These balances are the most important assumption of Chen's theory, because the internal energy exchanges are considered as external sources in the balances, that is, there is not associated entropy production (see e.g. [21]). Finally, observe that in Chen's approximation the total internal energy is conserved.

Following the usual procedure [25,26], the second law of thermodynamics

$$\frac{\partial s}{\partial t} + \nabla \cdot J_s \geq 0 \qquad (3)$$

leads to the constitutive equations for the ballistic and diffusive heat fluxes. In the above expression $s$ is the entropy density and $J_s$ is the entropy density flux. In order to get the constitutive equations we express $s$ and $J_s$ in the following way

$$s(e_b, e_d, q_b, q_d) = s_{eq}(e_b, e_d) - \frac{m_1}{2} q_d \cdot q_d - \frac{m_2}{2} q_b \cdot q_b \tag{4}$$

and

$$J_s = b_d \cdot q_d + b_b \cdot q_b, \tag{5}$$

with $m_1$ and $m_2$ positive constants and $b_d$ and $b_b$ second order tensors. These tensors are called *current multipliers* and they are interpreted as constitutive functions. By substituting Eqs. (4) and (5) in the second law, inequality Eq. (3), and introducing the internal energy balances as constrains we arrive to

$$\left(b_d - \frac{\partial s_{eq}}{\partial e_d} I\right) : \nabla q_d + \left(b_b - \frac{\partial s_{eq}}{\partial e_b} I\right) : \nabla q_b + \left(\nabla \cdot b_d - m_1 \frac{\partial q_d}{\partial t}\right) \cdot q_d + \left(\nabla \cdot b_b - m_2 \frac{\partial q_b}{\partial t}\right) \cdot q_b$$

$$+ Q\left(\frac{1}{T_d} - \frac{1}{T_b}\right) \geq 0,$$

(6)

where $I$ is the identity tensor. One can realize that the second part of the heat exchange, which is linear in the temperature does not contribute to the entropy production. This is an exact realization of the heat supply see e.g. Wang and Hutter [21]. If the material is one dimensional and isotropic, the linear solution of Eq. (6) is

$$b_d - \frac{1}{T_d} = k_{12} \frac{\partial q_b}{\partial x} + k_1 \frac{\partial q_d}{\partial x},$$

$$b_b - \frac{1}{T_b} = k_2 \frac{\partial q_b}{\partial x} + k_{21} \frac{\partial q_d}{\partial x},$$

$$\frac{\partial b_d}{\partial x} - m_1 \frac{\partial q_d}{\partial t} = l_1 q_d + l_{12} q_b,$$

$$\frac{\partial b_b}{\partial x} - m_2 \frac{\partial q_b}{\partial t} = l_{21} q_d + l_2 q_b,$$

$$Q = L\left(\frac{1}{T_d} - \frac{1}{T_b}\right) \tag{7}$$

where $\frac{\partial s_{eq}}{\partial e_b}$ and $\frac{\partial s_{eq}}{\partial e_d}$ have been interpreted as the inverse of the quasi-temperatures $T_b$ and $T_d$ respectively, and the following conditions must be satisfied by the coefficients $l_1$, $l_2$, $k_1$, $k_2$, $k_{12}$, $k_{21}$: $l_1 \geq 0$, $l_2 \geq 0$, $k_1 \geq 0$, $k_2 \geq 0$, $L \geq 0$, $k_1 k_2 - (k_{12} + k_{21})/4 \geq 0$, that is, the symmetric part of the conductivity matrix must be positive semidefinite, in order to fulfill the inequality of the second law, Eq. (6). In a three dimensional treatment one need a complete isotropic representation of the corresponding tensors. The above equations are simplified also by neglecting the cross coefficients in the last two equations, because these are not necessary for the reconstruction of the ballistic-diffusive theory. By eliminating the current multipliers from Eqs. (7), constitutive equations for the ballistic and diffusive heat fluxes as well as the are obtained:

$$\tau_d \frac{\partial q_d}{\partial t} + q_d + \ell_d q_b = -\lambda_d \frac{\partial T_d}{\partial x} + l_d^2 \frac{\partial^2 q_d}{\partial x^2} + \kappa_{12} \frac{\partial^2 q_b}{\partial x^2}, \tag{8}$$

$$\tau_b \frac{\partial q_b}{\partial t} + q_b + \ell_b q_d = -\lambda_b \frac{\partial T_b}{\partial x} + l_b^2 \frac{\partial^2 q_b}{\partial x^2} + \kappa_{21} \frac{\partial^2 q_d}{\partial x^2}, \tag{9}$$

$$Q = \Lambda(T_b - T_d), \tag{10}$$

respectively. In these equations $\tau_d$ and $\tau_b$ are relaxation times of the heat fluxes, $\lambda_d$, $\lambda_b$, $\kappa_{12}$ and $\kappa_{21}$ are material properties and $l_d$ and $l_b$ are interpreted here as the mean attenuation lengths of the heat carriers. $\ell_d$ and $\ell_b$ are coupling parameters. They are given in terms of the coefficients $l_1$, $l_2$, $k_1$, $k_2$, $k_{12}$, $k_{21}$ as follows: $\tau_d = m_1/l_1$, $\ell_d = l_{12}/l_1$, $\ell_b = l_{21}/l_2$, $\tau_b = m_2/l_2$, $\lambda_d = 1/(T_d^2 l_1)$, $\lambda_b = 1/(T_b^2 l_2)$, $\kappa_{12} = k_{12}/k_1$, $\kappa_{21} = k_{21}/k_2$, $l_d^2 = k_1/l_1$, $l_b^2 = k_2/l_2$, $\Lambda = L/(T_b T_d)$. The attenuation length interpretation is supported by dimensional considerations and also by the thermodynamic inequalities.

It should be noted that Eqs. (8) and (9) are coupled by cross-effects (terms in $\ell_d$, $\ell_b$, $\kappa_{12}$ and $\kappa_{21}$).

Then assuming that the cross effects associated to the Laplacian of the heat fluxes are negligible, that is $\kappa_{12} = \kappa_{21} = 0$, the combination of the constitutive equations with those of internal energy balance (with $\Lambda = 0$) is straightforward to eliminate the heat fluxes. Then, two coupled transport equations are obtained which, in dimensionless form, read

$$\alpha^2 \frac{\partial^2 T_d}{\partial t^2} + \left(\alpha - \frac{A\tau_d^2}{\tau_b C_d}\right)\frac{\partial T_d}{\partial t} - \frac{A\tau_d}{C_d}T_d + \ell_d \alpha \beta \frac{\partial T_b}{\partial t} + \frac{\ell_d A \tau_d}{C_d} T_b$$
$$= \left(\frac{1}{3} - \frac{A\tau_d}{C_d}\right) K_{nd}^2 \frac{\partial^2 T_d}{\partial x^2} + \alpha K_{nd}^2 \frac{\partial^3 T_d}{\partial t \partial x^2} \tag{11}$$

$$\frac{\partial^2 T_b}{\partial t^2} + \left(1 + \frac{A\tau_b}{C_b}\right)\frac{\partial T_b}{\partial t} + \frac{A\tau_b}{C_b}T_b + \frac{\ell_b}{\beta}\frac{\partial T_d}{\partial t} - \frac{\ell_b A \tau_b}{C_b} T_d$$
$$= \left(\frac{1}{3} + \frac{A\tau_b}{C_b}\right) K_{nb}^2 \frac{\partial^2 T_b}{\partial x^2} + K_{nb}^2 \frac{\partial^3 T_b}{\partial t \partial x^2} \tag{12}$$

where the timescale was fixed by $\tau_b$ and the length scale by an external length $L$. Equations (11) and (12) are the main result of this subsection. They are parabolic type equations due to the presence of terms in third order derivatives. The non-dimensional coefficients are given by the expressions

$$\alpha = \frac{\tau_d}{\tau_b}, \quad \beta = \frac{C_b}{C_d}, \quad K_{nd}^2 = 3\frac{\lambda_d \tau_d}{C_d L^2} = \frac{l_d^2}{L^2}, \quad K_{nb}^2 = 3\frac{\lambda_b \tau_b}{C_b L^2} = \frac{l_b^2}{L^2}. \tag{13}$$

It is interesting to note that these phenomenological coefficients contain information, on the one hand, on the thermal properties of materials and on the other, on properties of the interaction of the heat carriers with the lattice ($\alpha$ and diffusive and ballistic Knudsen numbers $K_{nd}$ and $K_{nb}$, respectively). We close this subsection by mentioning that if one assumes that $A = \frac{C_b}{\tau_b}$, as before to obtain Chen's internal energy balances and $\ell_b = \ell_d = 0$ (other couplings neglected), the following diffusive-ballistic transport equations are obtained,

$$\alpha^2 \frac{\partial^2 T_d}{\partial t^2} + \alpha(1-\alpha\beta)\frac{\partial T_d}{\partial t} + \alpha\beta T_d = \left(\frac{1}{3} - \alpha\beta\right)K_{nd}^2 \frac{\partial^2 T_d}{\partial x^2} + \alpha K_{nd}^2 \frac{\partial^3 T_d}{\partial t \partial x^2}, \tag{14}$$

$$\frac{\partial^2 T_b}{\partial t^2} + 2\frac{\partial T_b}{\partial t} + T_b = \frac{4}{3}K_{nb}^2 \frac{\partial^2 T_b}{\partial x^2} + K_{nb}^2 \frac{\partial^3 T_b}{\partial t \partial x^2}. \tag{15}$$

These equations may be compared with the diffusive-ballistic model by Lebon and Machrafi [20]. The second one coincides with Lebon's ballistic transport up to constants and the first one coincides with Lebon's diffusive transport equation up to constants and a coupling term which has been neglected here. Lebon and Machrafi's model will be compared with that obtained in this paper in a further subsection.

### 2.2 Effective medium theory

It is worth to mention that in Knudsen's numbers (Eq. 13), information about the microstructure of the superlattice may be introduced through the mean attenuation lengths of the heat carriers, which depends on the period thickness. Clearly, the coefficient $\alpha$ contains information about the microstructure since the relaxation times of the species also depend on the period thickness. The way in which the mean attenuation lengths and relaxation times depend on the number of periods can be obtained from a microscopic theory or from experimental measurements. This issue will be discussed in following sections.

Let us start by obtaining the dispersion relationships of the ballistic and diffusive heat carriers. This step implies a precise definition of the physical nature of the heat carrier. Let us consider that the energy is transported by traveling thermal waves of small amplitude $\delta T_0$, frequency $\omega$ and wave number $k$

$$T(x,t) = \delta T_0 e^{-i(kx-\omega t)}. \tag{16}$$

Substituting this expression in the transport Eqs. (11) and (12), the dispersion relationships are obtained. These rather long expressions will not be shown at this stage. Let us define now the effective thermal conductivity as

$$K_{eff} = -\frac{i\omega}{k^2}. \tag{17}$$

This definition is motivated by the dispersion relation of the Fourier theory. It defines the "modes" of heat conduction, like the propagation speeds defines the modes of wave propagation.

We assume that expression (17) describes the properties of heat transport in the superlattice. The effective thermal conductivities for the ballistic and diffusive carriers are obtained by using the dispersion relations leaving the result

$$K_{effd} = \frac{\alpha\left(1+\ell_d\beta-\frac{A\tau_d}{C_d}\right)+K_{nd}^2k^2}{2\alpha^2k^2}$$

$$\pm\frac{\sqrt{\left[\alpha\left(1+\ell_d\beta-\frac{A\tau_d}{C_d}\right)+K_{nd}^2k^2\right]^2-4\alpha^2\left[\frac{A\tau_d}{C_d}(\ell_d-1)+\left(\frac{1}{3}+\frac{A\tau_d}{C_d}\right)K_{nd}^2k^2\right]}}{2\alpha^2k^2},\qquad(18)$$

$$K_{effb} = \frac{1+\frac{A\tau_d}{C_d}+\frac{\ell_b}{\beta}+K_{nb}^2k^2}{2k^2}$$

$$\pm\frac{\sqrt{\left(1+\frac{A\tau_d}{C_d}+\frac{\ell_b}{\beta}+K_{nb}^2k^2\right)^2-4\left[(1-\ell_b)\frac{A\tau_b}{C_b}+\left(\frac{1}{3}+\frac{A\tau_b}{C_b}\right)K_{nb}^2k^2\right]}}{2k^2},\qquad(19)$$

respectively. If the two heat transport channels (ballistic and diffusive) are considered to act in parallel, the total effective thermal conductivity of the superlattice is given by:

$$K_{eff}^{SL} = K_{effb} + K_{effd}.\qquad(20)$$

This is the main result of this work. Expression (20) with (18) and (19) will be compared with some other formulations. In particular, in Subsection 4.1, it will be compared with that obtained from the high order dissipative fluxes formalism [22] and in Subsection 4.2 with Lebon-Machrafi's model [20]. In order to make contact with Lebon-Machrafi's theory we take $A = \frac{C_b}{\tau_b}$ to introduce Chen's balances in Eqs. (18,19). In such case the ballistic and diffusive effective conductivities reduce to the following expressions (we also take $\ell_b = \ell_d = 0$):

$$K_{effd} = \frac{-\alpha^2\beta+\alpha(1+K_{nd}^2k^2)}{2\alpha^2k^2} \pm \frac{\sqrt{(-\alpha^2\beta+\alpha(1+K_{nd}^2k^2))^2+4\alpha^2(\alpha\beta-\frac{1}{3})K_{nd}^2k^2+4\alpha^3\beta}}{2\alpha^2k^2},$$

(21)

$$K_{effb} = \frac{2+K_{nb}^2k^2}{2k^2} \pm \frac{\sqrt{(2+K_{nb}^2k^2)^2-4(1+\frac{4}{3}K_{nb}^2k^2)}}{2k^2},\qquad(22)$$

Expression (20) with (21) and (22) will be compared with the mentioned two formalisms in the following sections.

### 3. Results

*3.1. Heat carrier properties*

Let us consider a scenario which is consistent with the kinetic vision of the movement of heat carriers. In short, suppose that the ballistic carriers can move in each layer without interacting with the material but only with the interfaces that separate it from the rest of the system. Thus, the average length of attenuation can be expressed in terms of the thickness of the period $p$ in the superlattice as $l_b \approx p$. On the other hand, suppose that diffusive carriers have on average a certain attenuation length $l_d \ll L$. In these conditions, the Knudsen numbers can be expressed as

$$K_{nb} = 1, \qquad K_{nd} = \frac{l_d}{p}, \qquad (23)$$

where we have taken the period thickness as the characteristic length of the superlattice. The effective heat conductivities, as functions of the number of periods and wave number, are obtained by substituting the above expressions in Eqs. (18) and (19). They are used to calculate the effective heat conductivity of the superlattice accordingly with Eq. (20). We do this in the next subsection.

*3.2. Minimum effective heat conductivity*

In this subsection Eq. (20) with Eqs. (21) and (22) is fitted to the experimental values measured by Saha and collaborators [9,10]. To achieve this, one has two fitting parameters, namely, the mean attenuation length of the diffusive heat carriers and their wave number. Fig. 1 shows the result when $l_d = 25nm$ and $k = 1.2/m$.

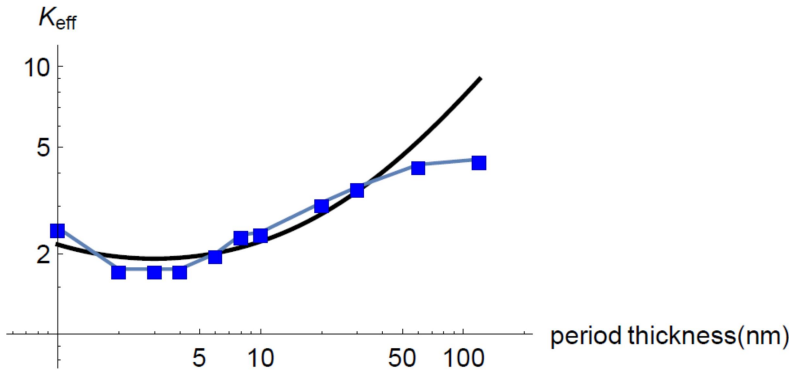

**Figure 1.** Effective thermal conductivity of the superlattice vs. period thickness. The model developed in this report predicts the experimentally found minimum around a period thickness of 4 nm. Black solid line: theoretical values; blue dots: experimental values from Saha et al. [9].

As it may be observed the effective thermal conductivity given by Eq. (20) shows a minimum in the effective heat conductivity around $p_{cr} = 4nm$ and the corresponding qualitative behaviour for $p < p_{cr}$ and $p > p_{cr}$.

## 4. Discussion

*4.1. Preliminar*

To summarize we remark two facts. First, we have obtained a heat transport model to describe effective properties of a superlattice from non-equilibrium thermodynamics with two internal variables, namely, the ballistic and the diffusive heat fluxes. Second, the derived equation (20), with (21) and (22), reproduces the minimum of the effective thermal conductivity in the superlattice. Our result shows a reasonable fitting with the experimental data- The appearance of the minimum is due to the presence of the third order terms in the derivatives in the transport equations (11) and (12).

Now we make some comparisons with other ballistic-diffusive formalism from the literature. In the study by Lebon et al. [19] two transport equations for the quasi-temperatures of ballistic and diffusive heat carriers were obtained under the assumption that the ballistic heat flux satisfies a Cattaneo constitutive equation and the diffusive heat flux a Guyer-Krumhansl equation. This differs from the scheme derived in this work wherein both of the fluxes satisfy Guyer-Krumhansl type equations. Our results show that this is crucial to understand the appearance of the minimum in the effective heat conductivity of the superlattice. Moreover, their heat transport equations are coupled by a term in the equation for the diffusive quasi-temperature which is linear in the ballistic quasi-temperature. The transport equation for the ballistic quasi-temperature does not contain coupling terms. In our formalism, an additional coupling term proportional to the time derivative of the ballistic quasi-temperature was obtained in the diffusive equation. It is worth mentioning that both terms come from a bit more general representation expressions of vector fields, Eq. (7). In obtaining the heat transport equations we have neglected the source term $Q$ in the balances equations [1,2]. The contribution of this term to the coupling problem deserves to be investigated. Our internal energy balance equations [1,2] contain Chen's theory as an approximation.

We finish the discussion with two short subsections where predictions of the effective heat conductivity from the high order dissipative fluxes formalism [22] and the ballistic-diffusive model by Lebon and Machrafi [19] are considered.

*4.2. High order dissipative fluxes formalism.*

In this subsection we compare the result of the previous section concerning the total effective heat conductivity to those predicted by the high order dissipative fluxes version of extended irreversible thermodynamics for a rigid solid by Álvarez and Jou [22]. In this formalism the generalized thermal conductivity is obtained as

$$K(k,\omega) = \frac{K^b(T)}{1 + i\omega\tau_1 + \dfrac{k^2 l_1^2}{1 + i\omega\tau_2 + \ldots}} \qquad (24)$$

being $K^b(T)$ the bulk thermal conductivity. The coefficients $\tau_i$ and $l_i$ have been interpreted as characteristic relaxation times and lengths of heat carriers, respectively. If all the relaxation times and characteristic lengths are considered to be equal, the asymptotic expression

$$K(T,k,\omega,l) = K^b(T) \left( \frac{-(1+i\omega\tau) + \sqrt{(1+i\omega\tau)^2 + (lk)^2}}{\dfrac{1}{2}(lk)^2} \right) \qquad (25)$$

is obtained from the continued fraction expression (24). This spectral expression establishes the contribution of propagating modes of wave number $k$ and frequency $\omega$ to the total thermal conductivity.

From Eq. (25) the thermal conductivity can be obtained in terms of the size $L$ of the system by taking $k = 2\pi/L$, which in the steady state takes the form:

$$K(T,l) = \frac{K^b(T)L^2}{2\pi^2 l^2}\left(\sqrt{1 + 4\left(\frac{\pi l}{L}\right)^2} - 1\right). \qquad (26)$$

$l$ is the mean attenuation length of heat carriers in the material. Equation (26) should be understood as the contribution of just the propagating mode with wave number $k = 2\pi/L$ to the heat conductivity of the system. Nevertheless, it has been useful to describe size effects in nanoscaled systems and reproduces satisfactorily experimental results [22]. Thus, the equation for the heat flux can be written as follows

$$q(x,t) = -K(T,l)\nabla T . \qquad (27)$$

This is a Fourier-like heat conduction law but it differs from the classical Fourier law in a fundamental way. It includes (through the generalized heat conductivity Eq. (24)) the dynamic effects of the complete hierarchy of high order dissipative fluxes. Equation (26) corresponds to the stationary state for particular values of the wave number. In this way, size effects have been incorporated to the effective heat conductivity. At this stage it may be assumed that the total heat flux decomposes into a ballistic and a diffusive part [22,24].

Now we proceed further by introducing the expression for the mean attenuation length of the diffusive heat carriers in Eq. (26) used in Subsection 3.1 (in order to fit our model to Saha`s experimental results) and take the characteristic length $L$ as the period thickness.

In Fig 2. it can be seen a plot of the effective thermal conductivity vs. period thickness as predicted by Eq. (26). In general the agreement of the calculated with the experimental data is good but it does not show the minimum in the effective thermal conductivity.

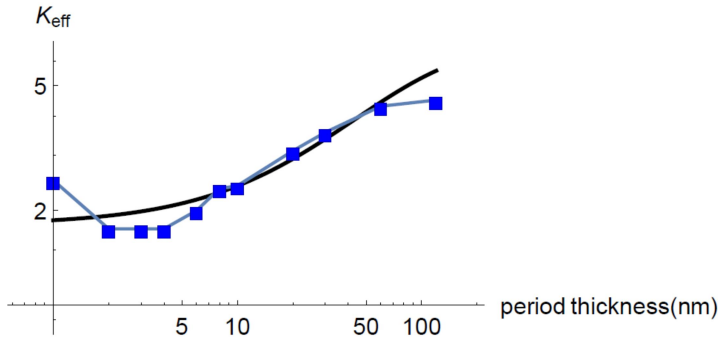

**Figure 2.** Effective thermal conductivity of the superlattice vs. period thickness according with the high order dissipative fluxes formulation of extended irreversible thermodynamics. This model does not predict the experimentally found minimum. Black solid line: theoretical values; blue dots: experimental values from Saha et al. [9].

*4.2. Lebon and Machrafi model [20]*

In this subsection we obtain the effective thermal conductivity from ballistic-diffusive model by Lebon et al. [19]. As mentioned, the main difference with the present formalism is that the ballistic quasi-temperature satisfies a Cattaneo equation while the diffusive quasi-temperature satisfies a Guyer-Krumhansl equation. We remind the reader that in the present formalism both, ballistic and diffusive quasi-temperatures, satisfy coupled Guyer-Krumhansl equations. We use the same scenario for the movement of the heat carriers described in Subsection 3.1. The result can be seen in Figure 3. It is observed that not only this model does not reproduce the minimum of the effective thermal conductivity but the fitting to the experimental values (blue dots) is not so good.

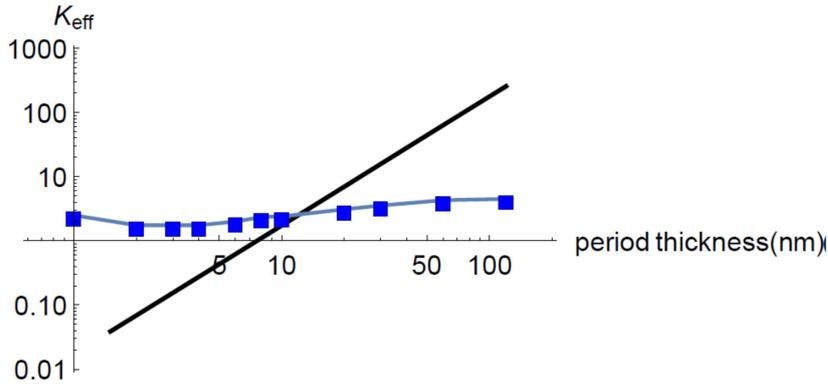

**Figure 3.** Effective thermal conductivity of the superlattice vs. period thickness from Lebon et al. model [19]. This model does not predict the experimentally found minimum. Black solid line: theoretical values; blue dots: experimental values from Saha et al. [9].

## 5. Conclusions

Some conclusions derived from the above study are in order.

The ballistic-diffusive model described in this work for the heat transport in superlattices contains the essence of the wave properties of heat carriers when heat waves are introduced in the formalism. The wave-like properties of the heat carriers are represented by the non-dimensional coefficients in Eqs. (11) and (12). The so obtained heat transport model reveals that the heat transport regime in the superlattice is composed by a part of heat carriers being diffusively transported and another part being ballistically transported through the system. The two regimes coexist and together constitute two heat transport channels with specific transport properties. The microstructure enters the formalism through the characteristic length of the superlattice, namely, the period thickness, and the mean attenuation length of heat carriers.

The developed effective ballistic-diffusive model requires that both components of the total heat flux satisfy Guyer-Krumhansl type equations which lead to coupled heat transport equations for the ballistic and diffusive quasi-temperatures. These equations are of the parabolic type and this seems to be the determining property to predict the existence of a minimum in the effective thermal conductivity as a function of the period thickness. Accordingly with the scenario depicted here in Subsection 3.1 for the properties of the heat carriers, the minimum is due to the fact that the long wavelength diffusive propagating modes have a mean attenuation length larger than the characteristic length of the superlattice and even larger than its total length.

The developed theoretical framework may shed light to the dependence of the mean attenuation length of the heat carriers on the microstructure. Through the proposed procedure to fit the model to the experimental data it can be found the explicit expression for the ballistic and diffusive Knudsen numbers as a function of the mean attenuation lengths. It would be expected that the prediction of the minimum of the effective thermal conductivity remain valid and a better fitting to the experimental data.

**Author Contributions:** Conceptualization, Federico Vázquez, Péter Ván and Róbert Kovács; Formal analysis, Federico Vázquez, Péter Ván and Róbert Kovács; Methodology, Federico Vázquez, Péter Ván and Róbert Kovács; Project administration, Róbert Kovács; Writing – original draft, Federico Vázquez; Writing – review & editing, Federico Vázquez and Péter Ván.

**Funding:** F.V. acknowledges financial support from CONACYT (México) reference CVU No. 10319, Tempus Public Foundation (Hungary) reference AK-00128-002/2018 and Energy Engineering Department of Budapest University of Technology and Economics (Hungary). The work was supported by the grants National Research, Development, and Innovation Office - NKFIH 124366(124508), 123815, KH130378, and FIEK-16-1-2016-0007.

**Acknowledgments:** We acknowledge Dr. B. Saha (Berkeley) and Dr. Y. Ezzahri (Poitiers) for interesting discussions on the problem treated in this paper. F.V. acknowledges the hospitality given by the Energy Engineering Department of Budapest University of Technology and Economics during his sabbatical stay.